\documentclass[usenatbib]{mn2e}
\usepackage{epsfig,rotating,natbib}

\def\br{\mathbf{r}}

\def\rh{q}
\def\ronh{q}

\newcommand{\pder}[2]{\ensuremath{\frac{\partial #1}{\partial #2}}}

\title[Adaptive softening with energy conservation]{An energy-conserving formalism for adaptive gravitational force softening in SPH and $N-$body codes}

\author[Price \& Monaghan]{D.J. Price$^{1}$, J.J. Monaghan$^{2}$ \\
$^1$School of Physics, University of Exeter, Exeter EX4 4QL, UK \\
$^2$School of Mathematical Sciences, Monash University, Clayton 3800, Australia\\
}
\pagerange{\pageref{firstpage}--\pageref{lastpage}} \pubyear{2006}
\date{Submitted: 23 September 2005}

\begin{document}
\label{firstpage}
\bibliographystyle{mn2e}
\maketitle

\begin{abstract}
 In this paper we describe an adaptive softening length formalism for collisionless $N-$body and self-gravitating Smoothed Particle Hydrodynamics (SPH) calculations which conserves momentum and energy exactly. This means that spatially variable softening lengths can be used in $N-$body calculations \emph{without} secular increases in energy. The formalism requires the calculation of a small additional term to the gravitational force related to the gradient of the softening length. The extra term is similar in form to the usual SPH pressure force (although opposite in direction) and is therefore straightforward to implement in any SPH code at almost no extra cost. For $N-$body codes some additional cost is involved as the formalism requires the computation of the density via a summation over neighbouring particles using the smoothing kernel.
 The results of numerical tests demonstrate that, for homogeneous mass distributions, the use of adaptive softening lengths gives a softening which is always close to the `optimal' choice of fixed softening parameter, removing the need for fine-tuning. For a heterogeneous mass distribution (as may be found in any large scale $N-$body simulation) we find that the errors on the least-dense component are lowered by an order of magnitude compared to the use of a fixed softening length tuned to the densest component. For SPH codes our method presents a natural and elegant choice of softening formalism which makes a small improvement to both the force resolution and the total energy conservation at almost zero additional cost.
\end{abstract}


\section{Introduction}
\label{sec:intro}
   This paper is concerned with the question of how best to represent the gravitational force when simulating self gravitating systems using particles. The simplest of such systems is a collection of stars which is usually replaced by a very much smaller number of computational particles.  A more complicated example is the simulation of self gravitating gas with or without a stellar component, using the particle method SPH \citep{monaghan05}.  
   
  Provided the number of computational particles is sufficient to resolve the important dynamical scales the simulation can give satisfactory results for most quantities though the slow relaxation of a galaxy is number dependent. In the case of $N-$body simulations it is, however, necessary to soften or smooth the forces between pairs of particles so that the binary collisions of the computational particles will not cause numerical artifacts.   
   
   The simple Plummer form of the softening where the force $F({\bf r})$ between a particle pair  with masses $m_a$ and $m_b$ separated by distance $\bf r$ is 
 \begin{equation}
  F({\bf r}) =  - G \frac{m_a m_b {\bf r} }{(r^2 + h^2)^{3/2}},
 \end{equation}
 and $h$ is the softening length. \citet{dehnen01}, amongst others \citep[e.g.][]{di93} has shown that a better choice is to use Kernel smoothing with kernels $W({\bf r},h)$  that have compact support. The softened force at $a$ due to $b$ then takes the form
\begin{equation}
 {\bf F} = - G \frac{4\pi m_a m_b {\bf r}}{r^3} \int_0^r W(r,h) r^2 dr.
\label{eq:F}
\end{equation}
 Provided $h$ is constant Poison's equation shows that the softening is equivalent to calculating the local gravitational force on a point particle $a$ due to a density
\begin{equation}
\rho({\bf r}_{a}) = m_b W({\bf r}_{a} - {\bf r}_b,h),
\end{equation}
or, when there is a collection of particles contributing to the force on particle $a$, the density is
\begin{equation}
\rho({\bf r}_{a}) = \sum_b m_b W({\bf r}_{a} - {\bf r}_b,h)
\end{equation}
 which is identical to the SPH density estimate.
 
 Kernels with compact support are zero beyond some specified distance proportional to the length scale $h$,  and the pair force then has the correct value for the two sets of real particles represented by two computational particles.
 
 The usual practice is to use a fixed value of $h$ for all of the $N-$body particles. A key issue which arises in this context, and which has been the subject of a number of studies \citep{merritt96,romeo98,athanassoula00,dehnen01,rs05}, is the `optimal' choice of softening length, for too small a softening length will result in noisy force estimates, whilst too large a value will systematically `bias' the force in an unphysical manner. In general, however, the `optimal' choice depends on particular system under investigation and may not be known \emph{a priori} \citep{athanassoula00}. \citet{dehnen01} quantifies the errors arising from both Plummer and kernel softening of the above form. In all cases he finds the accuracy is improved if  $h$ is allowed to vary according to the local particle number density $n$ in such a way that $h$ is smaller when $ n$ is larger.  Typically $h\propto 1/n^{1/3} $. For self-gravitating SPH calculations using a softening length which differs from the smoothing length can lead to unphysical results \citep{bateburkert97}. 
 
  To retain conservation of linear and angular momentum it is necessary to use a symmetric form of ${\bf F}$ so that each particle in a pair interaction experiences an equal but opposite force.  This can be achieved by using, for example $\bar h = \frac12(h_i +h_j) $ in place of  $h$ in (\ref{eq:F}).  However, because the softening length then varies in space the total energy of the system will not be conserved.  The errors are often not large but can lead to secular increases in the total energy of the system, destroying the phase-space conservation which is crucial for accurate $N-$body simulation \citep{hb90,dehnen01,rs05}. 
 
  In this paper we show how a Lagrangian for a self gravitating gas can be devised which has the softening of the force and the variation of $h$ built in.   The advantage of using a Lagrangian is that, provided it is constructed correctly, the conservation laws are automatically satisfied. In particular the conservation of energy and momentum is exact though, in practice, the accuracy is determined by the time stepping algorithm.  The new equations of motion have an extra term in addition to the standard SPH and gravity terms.  It is this term which guarantees energy conservation. We apply our algorithm to both static and dynamic problems. In some cases the new equations give results which are very similar to results obtained previously, but in some cases the results are very much improved. 
   
\section{Kernel Softening}
 A general formulation for force softening was given by \citet{dehnen01} and we use a similar formulation here. The modified gravitational potential per unit mass may be written in the form
\begin{equation}
\Phi({\bf r}) = -G \sum_{b=1}^{N} m_{b} \phi \left(\vert \br - \br_{b} \vert, h  \right)
\label{eq:Phi}
\end{equation}
where $\phi$ is the softening kernel which is a function of the particle separation and the softening length $h$ (we use $h$ to denote the softening length since it corresponds with the smoothing length used in the SPH density estimate). The kernel determines the functional form of the modified gravitational force law. For example, in the case of Plummer softening the softening kernel is given by
\begin{equation}
\phi(r,h) = \frac{1}{h}\left[1 + \left(\frac{r}{h}\right)^{2} \right]^{-1/2}.
\end{equation}

Neglecting the spatial variation of $h$ the force estimate corresponding to (\ref{eq:Phi}) is given by
\begin{equation}
\hat{\bf F}({\bf r}) = -\nabla \Phi = -G  \sum_{b=1}^{N} m_{b} \phi' \left(\vert \br - \br_{b} \vert, h  \right)\frac{\br - \br_{b}}{\vert \br - \br_{b} \vert},
\label{eq:fsoft}
\end{equation}
where $\phi' = \partial \phi / \partial \vert \br - \br_{b} \vert$. The underlying smooth density field can be obtained from Poisson's equation
\begin{equation}
\nabla^{2} \Phi = 4\pi G \rho,
\end{equation}
giving
\begin{equation}
\rho({\bf r}) = \sum_{b=1}^{N} m_{b} W\left(\vert \br - \br_{b} \vert, h  \right),
\label{eq:rho}
\end{equation}
where the density kernel is related to the softening kernel according to
\begin{equation}
W(r) = -\frac{1}{4\pi r^{2}} \pder{}{r} \left(r^{2} \pder{\phi}{r} \right).
\end{equation}
The kernel density given by (\ref{eq:rho}) corresponds to the mass distribution of each particle being smoothed. Readers familiar with SPH will notice that (\ref{eq:rho}) corresponds to the density estimate used in SPH calculations, where $W$ is the usual SPH smoothing kernel.
 
\begin{figure}
\begin{center}
\begin{turn}{270}\epsfig{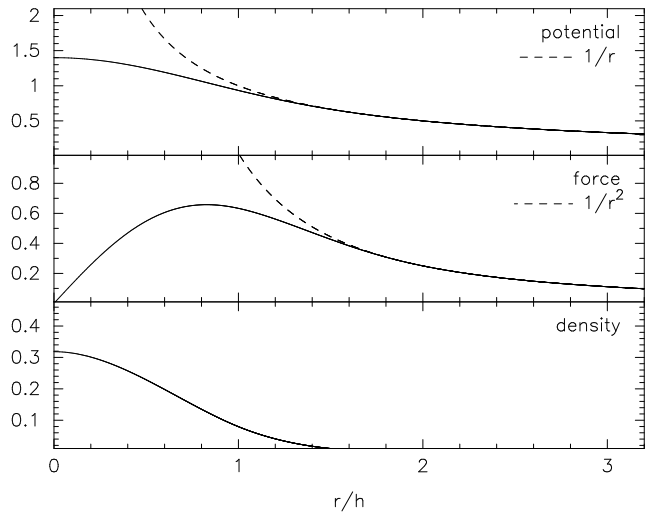}\end{turn}
\caption{The functional form of the modified potential (-), gravitational force and the density profile using the cubic spline kernel. For $r/h \ge 2$ the smoothing is zero and the potential and force are exact.}
\label{fig:kernels}
\end{center}
\end{figure}

 In general, the functional form of the softening kernel may be specified for either the potential term $\phi$, the force evaluation term $\phi'$ or $W$. In each case the corresponding kernel for the other cases may be determined by a straightforward integration or differentiation. For example, in $N$-body codes, it may be preferable to choose a kernel primarily for the force evaluation, from which the functional form of the potential and density kernel can be derived. In SPH the kernel is primarily used for the density estimate, where the most commonly used form is the cubic spline of \citet{ml85}
\begin{equation}
W(r,h) = \frac{1}{\pi h^{3}} \left\{ \begin{array}{ll}
1 - \frac32 \rh^2 + \frac34 \rh^3, & 0 \le \ronh < 1; \\
\frac14(2-\ronh)^3, & 1 \le \ronh < 2; \\
0. & \ronh \ge 2. \end{array} \right. \label{eq:cubicspline}
\end{equation}
where $q= r/h$. The corresponding force kernel is given by
\begin{equation}
\phi' = \frac{4\pi}{r^{2}} \int_{0}^{r} W r'^{2} dr',
\label{eq:phidash}
\end{equation}
the functional form of which is given for the cubic spline in Appendix~\ref{sec:cubicspline}.
The softening kernel for the gravitational potential may be calculated from the force kernel using
\begin{equation}
\phi = \int F dr,
\label{eq:Fdr}
\end{equation}
the form of which is also given in Appendix~\ref{sec:cubicspline} for the cubic spline. For general kernels (\ref{eq:Fdr}) combined with (\ref{eq:phidash}) can be integrated by parts to give
\begin{displaymath}
\phi(r,h) = 4\pi \left[ -\frac{1}{r} \int^{r}_{0} W r'^{2} dr' + \int^{r}_{0} W r' dr' - \int^{2h}_{0} W r' dr' \right],
\end{displaymath}
where the last term is the constant of integration, determined by the requirement that $\phi \to 0$ as $r\to\infty$ (note that we have also assumed a kernel with compact support of size $2h$).

 The modified potential, force functions and the density kernel are shown in Figure~\ref{fig:kernels} for the cubic spline. The reader should note that whilst we use the cubic spline as an example throughout this paper, the results derived in the following sections are quite general and any smoothing kernel may be used (including any of those suggested by \citealt{dehnen01}).
 
\section{Lagrangian formulation}
\label{sec:lagrangian}
The Lagrangian describing a self-gravitating gas is given by
\begin{equation}
L = \sum^{N}_{b=1} m_b \left(\frac12 {\bf v}_{b}^{2} - \Phi_{b} - u_{b} \right),
\label{eq:Lagrangian}
\end{equation}
where $\Phi$ is the gravitational potential (\ref{eq:Phi}) and $u$ is the thermal energy per unit mass. The equations of motion may be obtained via the Euler-Lagrange equations
\begin{equation}
\frac{d}{dt}\left(\pder{L}{{\bf v}_{a}} \right) - \pder{L}{{\bf r}_{a}} = 0,
\label{eq:el}
\end{equation}
giving
\begin{equation}
m_{a}\frac{d{\bf v}_{a}}{dt} = \pder{L}{{\bf r}_{a}}.
\end{equation}
 The advantage of using a Lagrangian to derive the equations of motion is that, provided the Lagrangian is symmetrised appropriately, momentum and energy conservation are guaranteed.
Variational principles have been used extensively to derive conservative SPH formalisms for relativistic fluid dynamics \citep{mp01}, magnetohydrodynamics \citep{pm04b} and in the case of a spatially variable smoothing length \citep{sh02,monaghan02}.

 An adaptive softening length formalism may be derived by writing the gravitational part of the Lagrangian in the form
\begin{eqnarray}
L_{grav} & = & -\sum_{b} m_{b} \Phi_{b}, \nonumber \\
& = & -\frac{G}{2} \sum_{b} \sum_{c} m_{b} m_{c} \phi_{bc}(h_{b}),
\label{eq:Lgravphiav}
\end{eqnarray} 
where $\phi_{bc}$ refers to $\phi(\vert \br_{b} - \br_{c} \vert)$. Swapping indices in the double summation shows that (\ref{eq:Lgravphiav}) is equivalent to averaging the softening kernels in the form
\begin{equation}
L_{grav} = -\frac{G}{2} \sum_{b} \sum_{c} m_{b} m_{c}\left[\frac{ \phi_{bc}(h_{b}) + \phi_{bc}(h_{c}) }{2}\right].
\end{equation}

The derivative of (\ref{eq:Lgravphiav}) is given by
\begin{eqnarray}
 \pder{L_{grav}}{{\bf r}_{a}} & = &  -\frac12 \sum_{b} \sum_{c} m_{b} m_{c} \left[ \left.\pder{\phi_{bc}(h_{b})}{\vert r_{bc} \vert}\right\vert_{h}  \pder{\vert r_{bc} \vert}{\bf r_{a}}  \right. \nonumber \\
&  & \hspace{25mm} + \left. \left.\pder{\phi_{bc}(h_{b})}{h_{b}}\right\vert_{r} \pder{h_{b}}{\br_{a}}\right],
\label{eq:sofar}
\end{eqnarray}
where
\begin{equation}
\pder{\vert r_{bc} \vert}{\bf r_{a}} = \frac{\br_{b} - \br_{c}}{\vert \br_{b} - \br_{c} \vert}(\delta_{ba} - \delta_{ca}).
\label{eq:drdr}
\end{equation}
We relate the smoothing length to the particle co-ordinates, assuming $h = h(\rho)$, using
\begin{equation}
\pder{h_{b}}{{\bf r}_{a}} = \pder{h_{b}}{\rho_{b}} \pder{\rho_{b}}{\br_{a}},
\label{eq:htox}
\end{equation}
where $\rho$ is the density calculated by a summation over neighbouring particles in the form
\begin{equation}
\rho_{a} = \sum_{b} m_{b} W\left(\vert \br_{a} - \br_{b} \vert, h_{a}\right),
\label{eq:rhosum}
\end{equation}
where $W$ is the density kernel. The relationship between $h$ and $\rho$ means that this is a non-linear equation for both $h_{a}$ and $\rho_{a}$ which can be solved self-consistently for each particle. The iterative method we use for doing so is described in detail in \S\ref{sec:settingh}. The spatial derivative of (\ref{eq:rhosum}) is
\begin{equation}
\pder{\rho_{b}}{{\bf r}_{a}} = \frac{1}{\Omega_{b}} \sum_{d} m_{d} \pder{W_{bd}(h_{b})}{{\bf r}_{a}} \left( \delta_{ba} - \delta_{da}\right),
\label{eq:gradrho}
\end{equation}
where $W$ is the density kernel and $\Omega$ is a term accounting for the gradient of the smoothing length given by
\begin{equation}
\Omega_a = \left[1 - \pder{h_a}{\rho_a}\sum_{b} m_{b}
\pder{W_{ab}(h_a)}{h_a}\right].
\label{eq:omega}
\end{equation}

 Using (\ref{eq:drdr}), (\ref{eq:htox}) and (\ref{eq:gradrho}) in (\ref{eq:sofar}) and simplifying, we have
\begin{eqnarray}
 \pder{L_{grav}}{{\bf r}_{a}} & = & -m_{a} \sum_{b} m_{b} \left[ \frac{\phi'_{ab}(h_{a}) + \phi'_{ab}(h_{b}) }{2}\right] \frac{\br_{a} - \br_{b}}{\vert \br_{a} - \br_{b}\vert} \\
 & & - m_{a} \sum_{b} m_{b} \frac12 \left(\frac{\zeta_{a}}{\Omega_{a}}  \pder{W_{ab}(h_{a})}{{\bf r}_{a}} + \frac{\zeta_{b}}{\Omega_{b}}  \pder{W_{ab}(h_{b})}{{\bf r}_{a}} \right). \nonumber
\end{eqnarray}
The quantity $\zeta$ is defined as
\begin{equation}
\zeta_{a} \equiv \pder{h_{a}}{\rho_{a}} \sum_{b} m_{b} \pder{\phi_{ab}(h_{a})}{h_{a}}.
\label{eq:gradsoftphiav}
\end{equation}
where $\partial \phi / \partial h$ can be tabulated (or calculated) directly for the particular smoothing kernel used. For the cubic spline the expression is given in Appendix~\ref{sec:cubicspline}. 

 The derivation of the SPH pressure force from the thermal energy term in the Lagrangian (\ref{eq:Lagrangian}) in the case of a spatially variable smoothing length has been described in detail elsewhere \citep[e.g.][]{sh02,monaghan02,pm04b} and we simply use the result here. The final equations of motion take the form
\begin{eqnarray}
\frac{d{\bf v}_{a}}{dt}  & = & -G\sum_{b} m_{b} \left[\frac{ \phi'_{ab}(h_{a}) + \phi'_{ab}(h_{b}) }{2}\right] \frac{\br_{a} - \br_{b}}{\vert \br_{a} - \br_{b} \vert} \label{eq:fgradsoftphiav}\\
& & -\frac{G}{2} \sum_{b} m_{b} \left[ \frac{\zeta_{a}}{\Omega_{a}} \pder{W_{ab} (h_{a})}{\br_{a}} + \frac{\zeta_{b}}{\Omega_{b}} \pder{W_{ab} (h_{b})}{\br_{a}}\right] \nonumber \\
& & -\sum_{b} m_{b}\left[ \frac{P_{a}}{\rho_{a}^2\Omega_{a}} \pder{W_{ab}(h_a)}{{\bf r}_a} + 
\frac{P_{b}}{\rho_{b}^2\Omega_{b}} \pder{W_{ab}(h_b)}{{\bf r}_a} \right]. \nonumber
\end{eqnarray}
 The first term in (\ref{eq:fgradsoftphiav}) corresponds to the softened gravitational force. The second term is present only in the case of adaptive softening lengths and it is the incorporation of this term which restores the energy conservation. The third term is the usual SPH pressure force allowing for a spatially variable smoothing length. The terms $\Omega$ and $\zeta$ required in the adaptive softening term (and for $\Omega$ also in the pressure force) are easily calculated alongside the density summation.
 
  The additional adaptive softening length term can be seen to have the same form as the pressure force, with the quantity $P/\rho^{2}$ replaced by $\zeta$. Notice however that $\zeta$ is, for positive kernels, a negative definite quantity and therefore that the adaptive softening term acts in opposition to the usual pressure term (that is, in the direction of increasing the gravitational force). This is in line with the recent findings of \citet{hubber06}; that SPH always \emph{underestimates} the gravitational force at low resolution. In fact they suggest adding an additional contribution to the gravitational force based on the `self-gravity' of an SPH particle. The new term derived above provides a similar contribution without the need for {\it ad-hoc} prescriptions.

 Alternative formulations of the adaptive softening formalism given above are possible by symmetrising the Lagrangian in different ways. We use the formulation given above since it is simple and efficient to implement. As an example, we derive an alternative version based on the average softening length in Appendix~\ref{sec:hav}. Whilst the force derived using the average softening length is very similar to  (\ref{eq:fgradsoftphiav}) but in keeping with the average $h$ used in the variable smoothing length SPH formalism of \citet{benz90}, it has the practical disadvantage that we do not use the average smoothing length elsewhere in the calculations, neither in the density summation (since the density for particle $a$ is calculated using only $h_{a}$) nor in the SPH force term (see \ref{eq:fgradsoftphiav}). Thus the calculation of the quantity $\bar{\zeta}$ (\ref{eq:gradsofthav}) for each particle requires the calculation of the kernel not only using $h_{a}$ (for the density and $\Omega$) but additionally using $\bar{h}_{ab}$ which is not only inefficient but also rather inelegant in the numerical code. Thus, we do not use the average $h$ formalism in this paper.
 
 A further possibility, not examined in this paper, would be to use the \emph{product} of the softening kernels in the force evaluation. The situation is complicated slightly in this case as the product form must either be used in the potential or force but not both. In any case the differences in the force evaluated using different symmetrised forms are very small. The suggestion put forward by \citet{di93} that the force should be symmetrised by considering two overlapping spheres may also be used in a similar manner to derive an energy conserving formalism, but it is not clear that there is any advantage to be gained by doing so \citep[see][]{dehnen01}.

 For reference the consistent forms of the continuity and internal energy equations for SPH simulations are given by
\begin{equation}
\frac{d\rho_a}{dt} = \frac{1}{\Omega_a} \sum_{b} m_{b} ({\bf v}_{a} - {\bf v}_{b})\cdot\pder{W_{ab}(h_a)}{{\bf r}_a},
\label{eq:sphcty}
\end{equation}
and
\begin{equation}
\frac{du_a}{dt} = \frac{P_a}{\Omega_a \rho_a^2} \sum_{b} m_{b} ({\bf v}_{a} - {\bf v}_{b})\cdot\pder{W_{ab}(h_a)}{{\bf r}_a},\label{eq:sphutherm}
\end{equation}
where ${\bf v}$ is the particle velocity. The continuity equation can be used to make a starting guess for the $h$ iteration procedure used to determine the density (described in the following section). An alternative to using the internal energy equation is to evolve the entropy as an independent variable \citep{sh02} which is possible for ideal equations of state.

\section{Numerical tests}
 We test the adaptive softening length formalism derived in the previous section using three examples. The first (\S\ref{sec:forces}) is a series of static tests used by \citet{dehnen01} and \citet{athanassoula00} in order to estimate the force errors associated with softening formulations. We also consider a dynamic version of one of these tests in order to study the energy conservation properties of our new method (\S\ref{sec:halorelax}). The second example (\S\ref{sec:poly}) involves self-gravitating SPH and the static structure and dynamical oscillation of a polytrope. 

\subsection{Errors}
 In the static halo tests, we calculate the Average Square Error (ASE) in the gravitational force according to
\begin{equation}
ASE = \frac{C}{N} \sum_{i=1}^{N} \vert f_{i} - f_{exact}({\bf x}_{i}) \vert^{2},
\end{equation}
where $f_{i}$ is the force on particle $i$, $N$ is the particle number and $C$ is a normalisation constant. Unless otherwise specified we use $C = 1/f_{max}^{2}$ where $f_{max}$ is the maximum value of the exact solution. The Mean Average Square Error (MASE) is then the mean over all realisations,
\begin{equation}
MASE = \frac{C}{N} \left \langle \sum_{i=1}^{N} \vert f_{i} - f_{exact}({\bf x}_{i}) \vert^{2}\right \rangle .
\end{equation}
 We choose this quantity rather than the Mean Integrated Square Error (MISE) used by \citet{merritt96} and \citet{dehnen01}, given by
\begin{equation}
MISE = \frac{C}{M}  \left \langle \int \rho({\bf x}) \vert f({\bf x}) - f_{exact}({\bf x}) \vert^{2} {\rm d}{\bf x}\right \rangle,
\end{equation}
where $\rho ({\bf x})$ is the true density at a point ${\bf x}$, $f({\bf x})$ is the force calculated at that point from the $N-$body distribution and $M$ is the total mass. Calculation of the MISE is complicated by the need to integrate along radial grid points (involving calculation of the force at positions other than particle positions) and \citet{athanassoula00} find little difference between their results using MASE or MISE error measures. In our case the correction terms derived in \S\ref{sec:lagrangian} depend on a particle's own density estimate, so it makes sense to calculate errors only at particle positions (that is, using the MASE estimate).

 The reader should bear in mind that, using either the MASE or MISE as defined above, the total error tends to be dominated by the regions containing the largest forces. This can be somewhat misleading in comparing adaptive softening with fixed softening, as the fixed softening length is generally chosen to minimise the error in the densest regions, where the adaptive softening will not show a large difference. An example is given in \S\ref{sec:bothplum} where a two-halo system is setup and we explicitly show the contribution to the MASE from each halo, even though the total error is dominated by the densest component.

\subsection{Setting the softening length}
\label{sec:settingh}
The method we use for setting the softening length is identical to the method used by \citet{price04} \citep[see][]{pm04b} for setting the smoothing length in SPH calculations. A similar method is also used by \citet{sh02} and hence also in the publicly available GADGET-2 code for $N$-body and SPH \citep{springel05}. The idea is to regard the smoothing length as a function of density via the relation
\begin{equation}
h_a \propto \rho_a^{-1/3},
\end{equation}
or more specifically
\begin{equation}
h_{a} = \eta \left(\frac{m_{a}}{\rho_{a}}\right)^{1/3},
\label{eq:hrho}
\end{equation}
where $m$ is the particle mass, $\rho$ is the mass density and $\eta$ is a dimensionless parameter which specifies the size of the smoothing length in terms of the average particle spacing (similar to the parameter $\epsilon$ used by \citealt{dehnen01}). The derivative is given by
\begin{equation}
\pder{h_{a}}{\rho_{a}} = -\frac{h_{a}}{3\rho_{a}}.
\end{equation}

An equivalent interpretation of (\ref{eq:hrho}) is that the mass contained within a smoothing sphere is held constant \citep{sh02}, that is
\begin{equation}
\frac{4}{3}\pi (\sigma h_{a})^{3} \rho_{a} = {\rm const} = m_{a} N_{neigh},
\label{eq:Nneigh}
\end{equation}
where $\sigma$ is the compact support radius of the kernel ($=2$ for the cubic spline) and $N_{neigh}\equiv\frac43\pi (\sigma \eta)^{3}$ may be used as an approximate measure of the number of neighbours contained within a smoothing sphere. Unless otherwise specified we use $\eta = 1.2$ in the variable smoothing/softening length formulations used throughout this paper, which in three dimensions is equivalent to $\sim 60$ neighbours.

 For a pure N-body simulation using unequal mass bodies, it may be advantageous to use a number density rather than the mass density for setting the softening length. The resulting gravitational force in that case is identical to the first two terms in (\ref{eq:fgradsoftphiav}) with mass density replaced by number density. In this paper we assume a mass density dependence consistent with SPH simulations.

  The density is calculated by a direct summation over the particles in the form (\ref{eq:rhosum}) which, via the relation (\ref{eq:hrho}) becomes a non-linear equation to be solved for both $h$ and $\rho$. \citet{dehnen01} suggests that even a rough approximation of the (number) density is sufficient for the purpose of adapting the softening length in $N-$body calculations. A similar argument may be made for setting the smoothing length in SPH calculations. In both cases, however, the situation changes once the gradient terms are incorporated into the equations of motion (as in this paper) since these terms are calculated on the basis of the $h(\rho)$ (or $h(n)$) relation and may therefore introduce substantial inaccuracies into the solution if the density and smoothing (or softening) length are far from being consistent with (\ref{eq:hrho}).

 From a practical point of view obtaining a self-consistent solution to (\ref{eq:hrho}) and (\ref{eq:rhosum}) is a relatively straightforward root-finding problem. The function to be solved may be written in terms of either $h$ or $\rho$. Written in terms of $h$ we have
\begin{equation}
f(h) = 0,
\end{equation}
in the form
\begin{equation}
\rho_{a}(h_{a}) - \rho_{sum}(h_{a}) = 0,
\label{eq:rhosolve}
\end{equation}
where $\rho_{a}$ is the density consistent with the current smoothing length $h_{a}$ calculated from the relation (\ref{eq:hrho}) and $\rho_{sum}$ is the density calculated using $h_{a}$ from the summation over neighbouring particles (\ref{eq:rhosum}). We use a Newton-Raphson iteration method, ie.
\begin{equation}
h_{a,new} = h_{a} - \frac{f(h_{a})}{f'(h_{a})} 
\end{equation}
where the derivative of (\ref{eq:rhosolve}) is given by
\begin{equation}
f'(h_{a}) = \frac{\partial \rho_{a}}{\partial h_{a}} - \sum_{b} m_{b} \frac{\partial W_{ab}(h_{a})}{\partial h_{a}} = -\frac{3\rho_{a}}{h_{a}}\Omega_{a}.
\end{equation}
We find this method to be efficient and cost effective, particularly since the quantity $\Omega$ (defined in (\ref{eq:omega})) is already calculated alongside the density summation for use in the equations of motion.

 Convergence is determined for each particle individually according to the criterion $\vert h_{new} - h \vert/h_{0} < \epsilon$ where $h_{0}$ is the smoothing (or softening) length at the start of the iteration procedure and typically we use $\epsilon = 10^{-3}$. We find that it is more efficient to perform the iterations by looping over the particles as the outer loop and iterating each particle at a time to convergence. We also find that it is no longer efficient to store a global neighbour list for all particles but rather to perform a neighbour search on-the-fly (e.g. using a treecode), recalculating where necessary and being stored only for the particle being iterated. This also represents a significant reduction in memory requirements for SPH calculations.
 
  The Newton-Raphson iterations work extremely well provided that the initial estimate of $h$ is reasonably close to the actual solution. This is almost always the case in the calculations since there is only a small change in $\rho$ between timesteps. The only problems which may arise are in the first iterations on the initial conditions where $h$ and $\rho$ may be far from the relation (\ref{eq:hrho}). For this reason it is useful to revert to a bisection scheme (which is guaranteed to converge) in the case where the Newton-Raphson iterations do not converge (we set the limit for this as $> 20$ iterations).

 In terms of cost the density iterations add only a small amount of extra work to SPH calculations. The exact work required depends on the nature of the simulation since more iterations are required when the density is changing rapidly. However the scheme is very efficient since iterations are only performed on the subset of particles whose densities are changing. The scheme can be made still more efficient by predicting an initial guess for $h$ in the time evolution scheme using the time derivative
\begin{equation}
\frac{dh}{dt} = \pder{h}{\rho} \frac{d\rho}{dt},
\end{equation}
where for the summation (\ref{eq:rhosum}) the time derivative is given by (\ref{eq:sphcty}). Using a prediction step we find that in general (although dependent on the dynamics of a particular simulation) only a small fraction of the particles require extra density calculations and that these particles then converge rapidly (in $\sim 2-3$ iterations).

\subsection{$N-$body tests}
\label{sec:forces}

\subsubsection{Isolated haloes}
\label{sec:static}
 The first test we perform is to compare the (softened) gravitational force to the exact force given an analytic density profile (corresponding to typical structures formed in cosmological simulations or used as initial conditions in galaxy models). Following \citet{dehnen01} we consider two different density profiles -- Plummer spheres and Hernquist models. The density profile for the Plummer spheres are given by 
\begin{equation}
\rho(r) = \frac{3GMr_{s}^{2}}{4\pi(r_{s}^{2} + r^{2})^{5/2}},
\end{equation}
where $M$ is the mass and $r_{s}$ is a parameter determining the concentration of the halo. The corresponding gravitational potential and force are given by
\begin{equation}
\Phi(r) = -\frac{GM}{(r_{s}^{2} + r^{2})^{1/2}}, 
\end{equation}
and
\begin{equation}
\Phi'(r) =  \frac{GM}{(r_{s}^{2} + r^{2})^{3/2}}.
\end{equation}
The cumulative mass profile for the Plummer sphere is given by
\begin{equation}
M(r) = \frac{GMr^{3}}{(r_{s}^{2} + r^{2})^{3/2}}.
\end{equation}

The \citet{hernquist90} model consists of a density profile given by
\begin{equation} 
\rho(r) = \frac{GMr_{s}}{2\pi r(r_{s} + r)^{3}}.
\end{equation}
 The density and gravitational forces in this model are more difficult to resolve since the density profile is cusped near the origin. The potential and force are given by
\begin{eqnarray}
\Phi(r) & = & -\frac{GM}{(r_{s} + r)}, \\
\Phi'(r) & = & \frac{GM}{(r_{s} + r)^{2}},
\end{eqnarray}
whilst the mass profile is given by
\begin{equation}
M(r) = \frac{GM r^{2}}{(r_{s} + r)^{2}}.
\end{equation}

The density profiles in each case are set up in the usual manner choosing three random deviates $(x_{1},x_{2},x_{3})$ uniformly on $(0,1)$. The first is used as a position in the mass profile from which the radial coordinate is determined by rearranging $M(r)$ to give $r(m)$ where $m$ is the mass fraction. In practice we use only mass fractions $< 0.99$ in order to prevent isolated particles being placed at extremely low densities. The second random number $x_{2}$ is used to give a random azimuthal angle $\varphi = \pi (2 x_{2} - 1)$ whilst the third is used to give a spherical angle $\theta$ via the transformation $\theta = \cos^{-1}{(2x_{3} - 1)}$ (necessary to prevent the distribution from clumping towards the poles). The result is a particle distribution which closely mirrors the analytic density profile although with errors decreasing like $1/\sqrt{N}$ due to the Monte Carlo nature of the distribution.
 
 In the numerical simulations we use units of mass $[M]=1$, length $[R]=1$ and time $[\tau] = (GM/R^{3})^{-1/2}$.  In these units $GM=1$ such that the gravitational constant does not appear in the numerical equations. Correspondingly, force and energy (both per unit mass) are measured in units of $GM/R^{2}$ and $GM/R$ respectively. For calculation of the mean error we compute $3\times 10^{6}/N$ realisations for a halo of $N$ particles.
  
\begin{figure}
\begin{center}

\begin{turn}{270}\epsfig{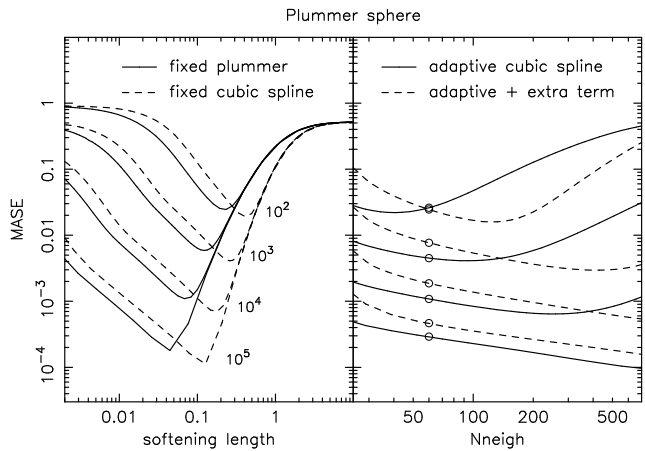}\end{turn}
\caption{MASE calculated for $3\times 10^{6}/N$ realisations of an isolated Plummer sphere with $M=1$, $r_{s}=1$ and $N=10^{2},10^{3}$, $10^{4}$ and $10^{5}$ particles. The left panel shows results using a fixed softening length, comparing Plummer softening (solid line) with cubic spline softening (dashed line). The right panel shows results using adaptive softening with (dashed line) and without (solid line) the new energy-conserving term. To guide the reader our fiducial choice of $N_{neigh} \simeq 60$ is indicated by the open circles.}
\label{fig:plummererrors}
\end{center}
\end{figure}

  The MASE calculated for Plummer haloes of $N=10^{2}$, $10^{3}$, $10^{4}$ and $10^{5}$ particles with $M=1$ and $r_{s} = 1$ are shown in Figure~\ref{fig:plummererrors}. The left panel shows the results using a fixed softening length, comparing both Plummer (solid lines) and cubic spline (dashed lines) softening kernels. The right panel shows the results using (cubic spline) adaptive softening, with (dashed line) and without (solid line) the energy-conserving term, also showing the variation with the ``Number of Neighbours'', by which we mean the parameter $N_{neigh}$ defined in (\ref{eq:Nneigh}).

\begin{figure}
\begin{center}

\begin{turn}{270}\epsfig{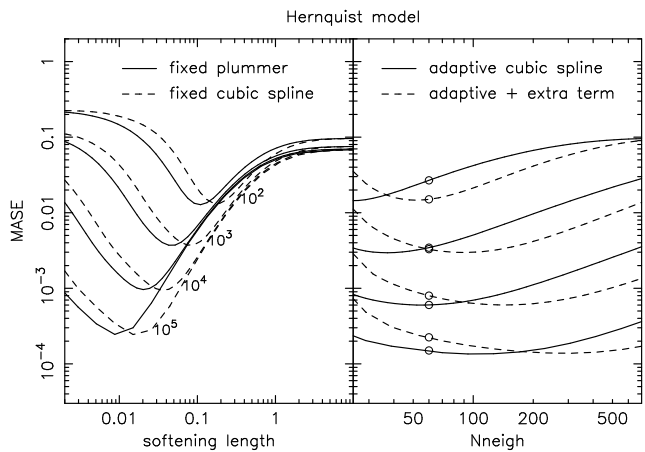}\end{turn}
\caption{As in Figure~\ref{fig:plummererrors} but for an isolated Hernquist model with $M=1$ and $r_{s}=1$. }
\label{fig:hernquisterrors}
\end{center}
\end{figure}

 Some general features are worth pointing out. Firstly, using a fixed softening length, there is a large variation in the total error depending on the choice of softening length (left panel). For softening lengths too small, the error is dominated by noise, reaching a maximum value once the softening length is smaller than the smallest particle separation. For softening lengths too large, the error is dominated by the \emph{bias} in the force introduced by the softening procedure. For some intermediate choice of softening length, there is a balance between noise and bias which produces a minimum error. This gives rise to the concept of `optimal softening' introduced by \citet{merritt96} and examined in detail by \citet{athanassoula00} and \citet{dehnen01}, whereby the softening length can be `fine tuned' for a particular simulation for minimum error. In principle this means that, for every $N-$body calculation, there is an optimal choice of softening length. The problem, demonstrated in the left panel of Figure~\ref{fig:plummererrors}, is that this `optimal' choice not only depends on the parameters of the problem (for example, Figure~\ref{fig:plummererrors} shows that the optimal choice changes with resolution, but \citet{athanassoula00} also discuss the dependence on the shape and degree of central concentration of the halo), but may also be different for different components of the same simulation. 
 
 By contrast, use of adaptive softening (right panel) shows only a weak dependence on the choice of the (adaptive) softening parameter (provided that the neighbour number is small compared to the total number of particles). To guide the reader, our fiducial choice of $N_{neigh}\simeq 60$ (given by $\eta = 1.2$ in (\ref{eq:hrho})) is marked by the open circle in each case. Making this reasonable choice in all cases gives a softening which (according to the MASE estimate) is close to the optimal choice of fixed softening. The exception is perhaps for the $10^{4}$ particle halo, where the optimal choice of fixed softening gives a MASE $\sim 2.5\times$ lower than that using adaptive softening with energy conservation. However, changing the fixed softening length up or down by a factor of two in either direction (ie. not a large range if the optimal choice is not known \emph{a priori}) means that even in this case that adaptive softening, even with the energy conservation term added, wins.
 
  Comparison of the dashed (Plummer kernel) and solid lines (cubic spline kernel) in the left panel of Figure~\ref{fig:plummererrors} confirms the conclusion reached by several authors that use of the cubic spline kernel is advantageous over the standard Plummer softening. In particular the optimal error for each halo is reached at a higher softening length for the cubic spline kernel and the slope in the error curve at high softening lengths is steepened, demonstrating that the cubic spline reduces the \emph{bias} in the force estimate. This is a result of the compact support of the cubic spline kernel, which gives a force with zero bias outside of the kernel radius (see discussion in \S\ref{sec:intro}).
 
  The right panel of Figure~\ref{fig:plummererrors} also shows the influence of the new energy-conserving term in (\ref{eq:fgradsoftphiav}) on the force errors in a static configuration. This term appears to increase the noise but also lower the bias in the (adaptively softened) force estimate, meaning that the total error is greater for smaller neighbour numbers but lower for larger neighbour numbers. We attribute this to the fact that the extra term is related to the gradients in softening length: Where these gradients are spurious (due to noise), the extra term may increase the total error. Where the gradients are due to actual gradients in the density, the extra term correspondingly leads to a more accurate force estimate. This conclusion is also borne out by the results using Hernquist models (right panel of Figure~\ref{fig:hernquisterrors}). Here the extra term leads to a smaller MASE (compare the dashed and solid lines in the right hand panel of Figure~\ref{fig:hernquisterrors}) at a lower $N_{neigh}$ than in the Plummer case (dashed vs solid lines in the right hand panel of Figure~\ref{fig:plummererrors}). The density profile in the Hernquist model is strongly cusped near the origin, meaning that any improvement in the resolution of density gradients (e.g. from the new term) tends to improve the error estimate.

 The Hernquist model was computed using $M=1$, $r_{s} = 1$ and $N=10^{2},10^{3}$, $10^{4}$ and $10^{5}$ particles. The results using fixed Plummer (solid lines) and cubic spline (dashed lines) softening on the Hernquist model are shown in the left hand panel of Figure~\ref{fig:hernquisterrors}. The differences between Plummer and cubic spline softening are much smaller in this case than for the Plummer spheres (Figure~\ref{fig:plummererrors}), apart from a factor of $\sim 2$ difference in the optimal choice of softening length for each kernel (that is, the optimal softening length for Plummer softening is approximately $\sim 1/2$ of the optimal value using cubic spline softening).

 These tests demonstrate that, for an isolated halo, the use of adaptive softening gives force errors which are close to optimal. Whilst there is not a significant improvement in the MASE compared to the use of an optimally-chosen fixed softening length, the use of adaptive softening removes the need for such fine-tuning. Furthermore it may not be possible to find a softening which is `optimal' for all components of a simulation. In the following section we consider such an example, where the use of adaptive softening shows a clear improvement.

\subsubsection{Two Plummer spheres}
\label{sec:bothplum}
 Next we consider two Plummer spheres placed at a fixed distance from each other, of equal mass but where one halo is much denser than the other. This situation may be representative of two haloes present in a typical cosmological $N-$body simulation or in a simulation of galaxy dynamics where more than one galaxy is present. A similar test was considered by \citet{athanassoula00} where a variety of mass ratios were also examined. Here we simply choose one representative case.
 
  Both haloes are Plummer spheres, setup as described in \S\ref{sec:static}. We use equal mass spheres with $M=0.5$. The first sphere is placed at the origin, with concentration parameter $r_{s} = 1$ whilst a second sphere with $r_{s} = 0.1$ (ie. much denser) is placed some distance away at $[x,y,z] = [10,0,0]$. The MASE is calculated un-normalised in this case, that is with the normalisation factor $C=1$ in order to make a meaningful comparison between the errors in each component.
  
\begin{figure}
\begin{center}
\begin{turn}{270}\epsfig{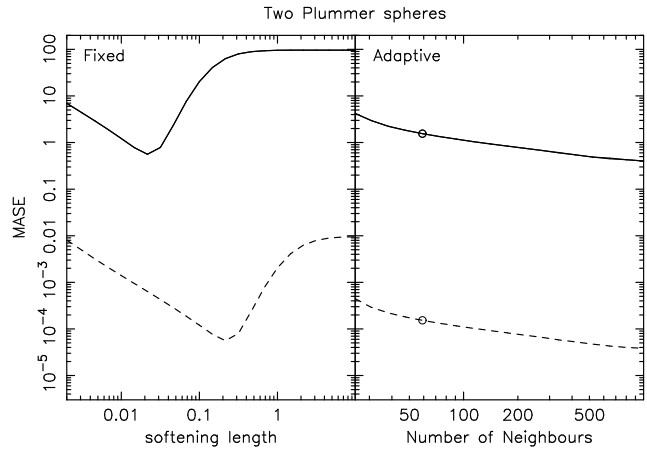}\end{turn}
\caption{Mean averaged square errors in the gravitational force calculated for 300 realisations of a configuration involving two Plummer spheres. The total MASE is given by the solid line whilst the contribution to the total MASE from the least dense component is given by the dashed line. Results using a fixed cubic spline softening, varying the softening length, are shown in the left panel. The right panel shows the results using our adaptive softening formalism (including the energy conservation term), varying the $N_{neigh}$ parameter. The open circles correspond to our fiducial choice of $N_{neigh} \simeq 60$.}
\label{fig:bothplum10k}
\end{center}
\end{figure}

   The MASE resulting from 300 realisations of this configuration using a total of 10,000 particles (5000 per sphere) is shown in Figure~\ref{fig:bothplum10k} (solid line), using fixed cubic spline softening (left panel) and adaptive softening with the energy conservation term included (right panel), showing the variation with softening length in the former case and $N_{neigh}$ in the latter (where again the open circles correspond to our fiducial choice of $N_{neigh}\simeq 60$). The total MASE (solid line) is completely dominated by the densest component, with results comparable to those shown in Figure~\ref{fig:plummererrors}. However, we also plot the contribution to the total MASE from the less-dense component (that is, the $r_{s}=1$ sphere at the origin) (dashed line).

 The problem with the use of a fixed softening length in a general $N-$body simulation is evident from Figure~\ref{fig:bothplum10k}, namely that the `optimal' choice of softening length differs for each component. The choice which minimises the errors in the densest component produces errors in the least dense component that are over $1\frac12$ orders of magnitude larger than the optimal choice of softening for that component. Conversely choosing the softening which is optimal for the least dense component produces a bias in the force estimate in the densest component leading to a MASE $\sim2$ orders of magnitude larger than the optimal choice of softening length in the dense component. Usual practise is therefore to choose the softening which minimises the softening in the densest component(s) (since this represents the largest contribution to the total error). However frequently one is interested in the properties of both (or all) components in an $N-$body simulation. This leads naturally to a need for adaptive force softening.
 
 The results using our adaptive softening length formalism (including the energy conservation term) are shown in the right hand panel of Figure~\ref{fig:bothplum10k}. For both components the resulting error is, as previously, close to the optimal choice of fixed softening, but here the softening is close to optimal for \emph{both} components. This means that the force errors in the least dense component are $\sim 1$ order of magnitude smaller than using a fixed softening length tuned to the densest component.

\subsubsection{Halo relaxation}
\label{sec:halorelax}
 An extension to the static halo test is to examine the \emph{dynamic} influence of the energy-conserving term. The initial conditions for this test are a Plummer sphere with an isotropic velocity distribution corresponding to a (dynamic) steady state. The particle velocities are setup in the manner described by \citet*{ahw74}: The distribution function is
\begin{equation}
f({\bf r}, {\bf v},0) = \left\{ \begin{array}{ll}
\frac{24\sqrt{2}}{7\pi^{3}} \frac{r_{s}^{2}}{G^{5}M^{4}}(-E)^{7/2} & E < 0, \\ 0 & E > 0,\end{array} \right.
\label{eq:distfn}
\end{equation} 
where $f({\bf r}, {\bf v},t) d{\bf r}d{\bf v}$ is the total mass of particles with position ${\bf r}$ and velocity ${\bf v}$ at time $t$ and $E$ is the energy per unit mass of a body:
\begin{equation}
E =\frac12 v^{2} + \Phi.
\label{eq:ener}
\end{equation}
 The distribution function is sampled by scaling the velocities in terms of the maximum velocity at $r$, ie. the escape velocity
\begin{equation}
v_{esc} = \sqrt{-2\Phi} = \sqrt{\frac{2GM}{ (r^{2} + r_{s}^{2})^{1/4}}}.
\label{eq:vesc}
\end{equation}
Writing $q=v/v_{esc}$, from (\ref{eq:ener}) and (\ref{eq:distfn}) the probability distribution for $q$ is proportional to
\begin{equation}
g(q) = q^{2} (1-q^{2}),
\end{equation}
where $\vert q \vert < 1$. This distribution is sampled using the Von Neumann rejection technique \citep{numericalrecipes}: two uniform random deviates $x_{4}$ and $x_{5}$ are drawn. Noting that $g(q)$ is always less than 0.1 (since $\vert q \vert < 1$) we adopt $q=x_{4}$ if $0.1x_{5} < g(q)$, otherwise a new pair of random numbers is tried until the inequality is satisfied. The velocity modulus $v$ is obtained using (\ref{eq:vesc}) and, using two more uniform random deviates $x_{6}$ and $x_{7}$, the velocities are given by
\begin{eqnarray}
v_{x} & = & (1 - 2x_{6})v, \nonumber \\
v_{y} & = & \sqrt{v^{2} - v_{x}^{2}} \cos{(2\pi x_{7})}, \nonumber \\
v_{z} & = & \sqrt{v^{2} - v_{x}^{2}} \sin{(2\pi x_{7})}.
\end{eqnarray}

 The halo is evolved forwards in time using a standard second order leapfrog integrator with a global timestep controlled by the condition
\begin{equation}
\Delta t = 0.15 \left(\frac{h}{f}\right)^{1/2}
\end{equation}
 where $h$ is the softening length, $f$ is the force per unit mass and the minimum over all particles is used.

\begin{figure}
\begin{center}
\begin{turn}{270}\epsfig{file=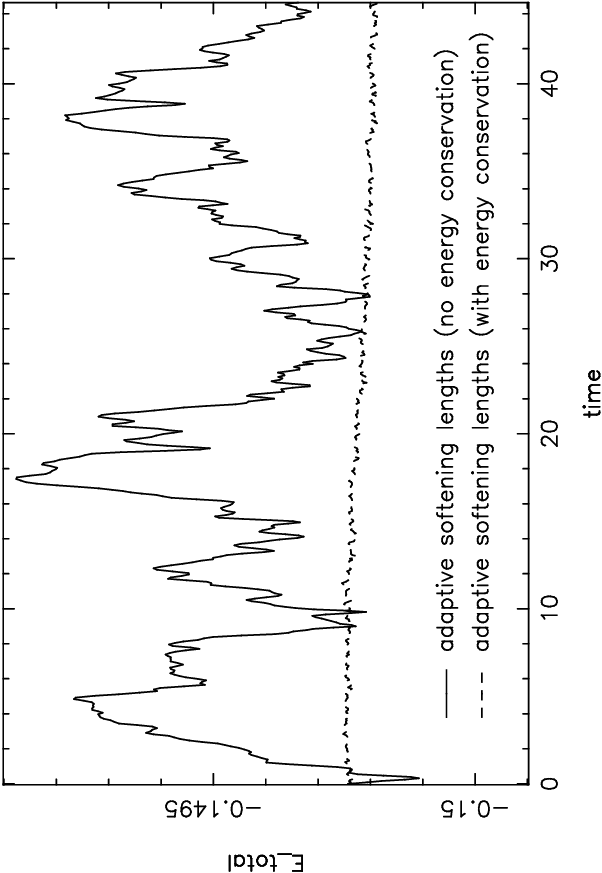, height=\columnwidth}\end{turn}
\caption{Total energy conservation during the dynamical evolution of the 1000-particle Plummer sphere. Using adaptive softening lengths without the additional term (solid line) leads to fluctuations in the total energy which dominate over the timestepping errors. Incorporating the new adaptive softening length term (dashed line), energy conservation is restored to timestepping accuracy.}
\label{fig:plummer_relax}

\begin{turn}{270}\epsfig{file=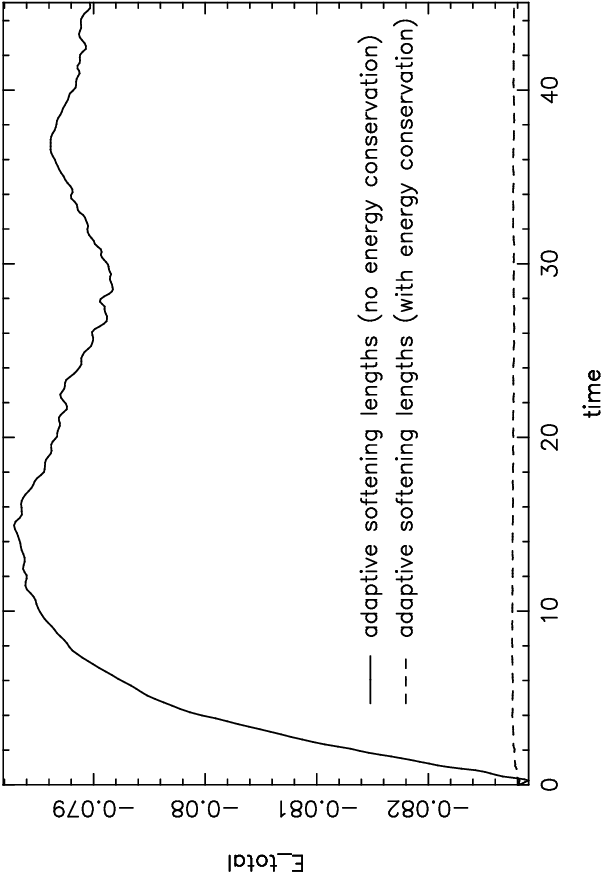, height=\columnwidth}\end{turn}
\caption{Total energy conservation during the dynamical relaxation of the perturbed 1000-particle Plummer sphere. In this case the initial velocities were multiplied by a factor of 1.2. Using adaptive softening lengths but without the new term (solid line) the change in the softening lengths caused by the initial expansion can be seen to cause a secular increase in the total energy. Adding this term (dashed line), the total energy is conserved to timestepping accuracy.}
\label{fig:plummer_expand}
\end{center}
\end{figure}

  The energy conservation during the evolution of the equilibrium halo is shown in Figure \ref{fig:plummer_relax}. Using adaptive softening without the additional term (solid line), fluctuations in the energy are observed from the changes in softening length which, although small, dominate over the errors due to timestepping. With the energy conserving term added (dashed line), only a small non-conservation of energy remains which can be shown to decrease as the timestep is made shorter. 
  
   As a slightly more demanding test, we also consider the relaxation of a perturbed Plummer sphere - that is, with the velocities drawn from the equilibrium distribution function as described above, but then multiplied by a factor of 1.2. This means that the halo initially expands before slowly relaxing into a dynamical equilibrium state. The evolution of the total energy in this case is shown in Figure~\ref{fig:plummer_expand}. Using adaptive softening lengths, the change in softening lengths corresponding to the initial expansion are reflected as a secular increase in the total energy (solid line). Using the new energy conserving formalism (dashed line), this secular increase is not present and total energy is conserved to timestepping accuracy.

\subsection{SPH tests}
\label{sec:poly}

\subsubsection{Static structure of a Polytrope}
 A simple test of self-gravitating gas dynamics is to verify the static structure of a polytrope by allowing an initial arrangement of gas to settle into hydrostatic equilibrium. In order to do so we setup $\sim 1000$ SPH particles in a quasi-uniform spherical distribution and damp them into an equilibrium state using a polytropic equation of state $P = K \rho^{\gamma}$ with $\gamma = 5/3$. The low resolution is chosen in order to highlight the differences between various softening formalisms.
 
  The exact manner in which the particles are initially setup is not particularly important, although a perfectly uniform arrangement tends to produce numerical artifacts in the collapsed particle configuration whilst a clumpy initial setup takes longer to settle to equilibrium. In this paper we use a quasi-uniform distribution achieved by placing particles initially on a uniform square lattice, cropped to ensure that $r < 1$ and with a small, random perturbation of amplitude $0.2\Delta$ (where $\Delta$ is the lattice spacing). The particle configuration is shifted slightly to ensure that the centre of mass is placed at the origin. Using a lattice spacing of $\Delta = 0.15$ results in a total of $1086$ particles in the calculations.
  
   The exact solution for the polytrope static structure is computed by solving the equation
\begin{equation}
\frac{\gamma K}{4\pi G (\gamma - 1)} \frac{d^2}{dr^2}
\left[r\rho^{\gamma-1}\right] + r\rho = 0,
\end{equation}
numerically using a simple finite difference scheme. The solution is then scaled to give a polytrope of radius unity. In code units (discussed in \S\ref{sec:forces}) a polytrope of radius unity is obtained by choosing $K = 0.4246$ in $P = K\rho^{\gamma}$.
  
 In all simulations the density and SPH smoothing length are calculated by direct summation using the iterative method described in \S\ref{sec:settingh}. Also the variable smoothing length terms in the SPH equations are used throughout. In order to isolate the effects of the softening formalisms we calculate the gravitational term by a direct summation over the particles (rather than using a treecode) with a standard second-order leapfrog scheme for time integration using a timestep controlled by a Courant condition based on the signal velocity \citep{monaghan05}. The particles are damped to an equilibrium using a standard form of the SPH artificial viscosity \citep{monaghan97} together with a damping in the force equation which is independent of particle number, given by
\begin{equation}
\frac{d{\bf v}}{dt} = - 0.05{\bf v} + {\bf f},
\end{equation}  
where ${\bf f}$ is the force per unit mass. Note that the polytropic equation of state means that the kinetic energy removed by the artificial viscosity and damping terms is not deposited as thermal energy but rather allowed to escape from the system.

\begin{figure}
\begin{center}
\begin{turn}{270}\epsfig{file=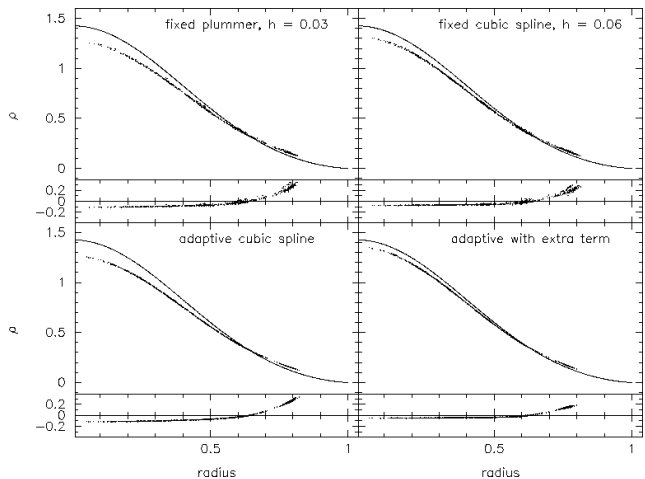, height=\columnwidth}\end{turn}
\caption{Static structure of the $\gamma = 5/3$ polytrope calculated using $1086$ SPH particles (solid points). The results are shown using fixed plummer softening with softening length $h=0.03$ (top left), fixed cubic spline softening with $h=0.06$ (top right), using adaptive softening lengths (bottom left) and finally using the energy-conserving formalism including the additional force term (bottom right). The exact solution is given by the solid line.  Note that the SPH smoothing length is adaptive in all cases}
\label{fig:staticpolytrope}
\end{center}
\end{figure}

 The equilibrium configuration of the $\gamma = 5/3$ polytrope with various softening formulations are shown in Figure~\ref{fig:staticpolytrope} and may be compared in each case to the exact solution given by the solid line. The fractional errors $(f_{i} - f_{exact})/ f_{exact}$ are also shown in an inset plot in each  panel. The top two panels show the results using fixed Plummer (top left) and cubic spline (top right) softening, where, not knowing the `optimal' choice \emph{a priori} we have used the rule-of-thumb given by \citet{springel05} whereby the softening length is chosen to be $\sim 1/40$ of the average particle spacing in the initial conditions. Thus guided we choose $h=0.06$ for the cubic spline softening, using half of this value, $h=0.03$, in the Plummer softening (see discussion in \S\ref{sec:static}).
 
 Using adaptive softening lengths without the energy conservation term (bottom left panel) shows a small improvement over the fixed softening results, mainly in the outer regions where the force estimate is much less noisy. The density resolution in the centre is slightly lower in this case, but this is substantially improved when the energy conserving term is incorporated (lower right panel). The error in the outer regions is also improved by the energy-conservation term. The more compact distribution produced in this case is consistent with the additional term being always in the direction of increasing the gravitational force (see \S\ref{sec:lagrangian}).

\subsubsection{Polytrope oscillations}
\begin{figure}
\begin{center}
\begin{turn}{270}\epsfig{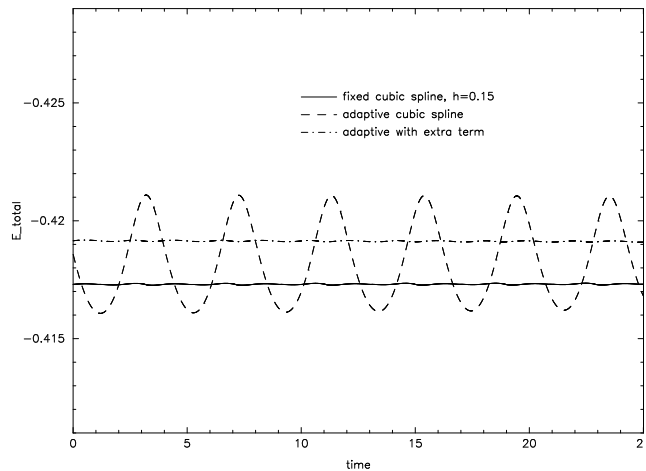}\end{turn}
\caption{Total energy conservation during the radial oscillations of the polytrope. The results are shown using a fixed softening length (solid line), adaptive softening lengths (dashed line) and using the new adaptive softening length formalism (dot-dashed line). Note the improvement in energy conservation in the adaptive softening case when the new term is included. The absolute value of the total energy differs slightly between runs because of the difference in equilibrium structure (Figure~\ref{fig:staticpolytrope}).}
\label{fig:polyoscills}
\end{center}
\end{figure}

 Having obtained the static structure, studying the radial oscillations of the polytrope provides a test of the energy conservation properties of the code. In order to do so we apply a radial velocity perturbation of $v_{r} = 0.2 r$ to the static solutions obtained in the previous section.
In order to distinguish effects due to the softening formulation from effects due to the timestepping algorithm we use a very low Courant number of $C_{cour} = 0.05$ for this test. In general, however, non-conservation effects from the softening formalism are much larger than effects due to timestepping. No artificial viscosity or damping is applied for this problem. 

The evolution of the total energy of the system is shown in Figure~\ref{fig:polyoscills} using cubic spline softening with fixed and adaptive softening lengths. Using a fixed softening length (solid line) the total energy is conserved exactly (ie. to timestepping accuracy). Adapting the softening length using the method described in \S\ref{sec:settingh} results in non-conservation of energy (dashed line). Incorporating the additional pseudo-pressure term into the adaptive softening formulation restores the total energy conservation (dot-dashed line).

\section{Summary}
 In this paper we have described an algorithm for using adaptive softening lengths in both SPH and N-body codes which retains the conservation of both momentum and energy. The formalism requires the computation of an additional gravitational force term which is similar in form to the SPH pressure force and is therefore straightforward to implement in any SPH code at almost no added cost. For pure N-body codes calculation of the additional term requires some extra work since quantities such as the density must be evaluated using the smoothing kernel. However even in this case the cost is small compared to the evaluation of the long-range gravitational forces using a treecode.

 The softened gravitational force can be symmetrised either by using an average of the softening lengths or alternatively an average of the softening kernels, where the latter is preferred because of the manner in which the density is calculated. The choice of softening kernel is completely arbitrary, with calculations in this paper were made using the standard SPH cubic spline kernel (although any of the kernels proposed by \citealt{dehnen01} could be used). 
 
 Use of spatially variable (`adaptive') softening lengths is found to provide near-optimal softening for arbitrary mass distributions using a single, fiducial choice of the adaptive softening parameter $N_{neigh}$. This contrasts to the results of \citet{athanassoula00} where the optimal (fixed) softening length was found to depend strongly on the number of particles and parameters such as the central concentration and shape of the mass distribution. For a mass distribution where more than one component is present, we find that the use of our adaptive softening length formalism can give more than an order of magnitude improvement in the errors on the least dense component.

 The main advantage of the formalism presented here is that adaptive softening lengths can be used whilst maintaining energy conservation to timestepping accuracy. This was found to be particularly important in the case of collisionless $N-$body simulations where secular increases in the total energy were found to result from the use of adaptive softening lengths without the energy-conserving term. For self-gravitating SPH simulations the new formalism is a natural and self-consistent choice which is found to give a small improvement in resolution and energy conservation over traditional \emph{ad-hoc} formulations for essentially zero additional cost.
  
\section*{Acknowledgements}
Many thanks go to Matthew Bate for many fruitful discussions and to the anonymous referee for comments which have improved this paper substantially.
DJP acknowledges the support of a PPARC postdoctoral fellowship and thanks Walter Dehnen for useful discussions. Computations were performed on the School of Physics iMac cluster at the University of Exeter.

\bibliography{sph,starformation}

\appendix
\begin{onecolumn}
\section{Cubic spline softening}
\label{sec:cubicspline}
 In this appendix we give the functional form of the softening corresponding to the cubic spline kernel  (\ref{eq:cubicspline}). Integrating the kernel according to (\ref{eq:phidash}), we find that the gravitational force is softened using
\begin{equation}
\phi'(r,h) = \left\{ \begin{array}{ll}
1/h^{2} \left[\frac43 \rh - \frac65 \rh^3 + \frac{1}{2}\rh^4 \right], & 0 \le \ronh < 1; \\
1/h^{2} \left[\frac83 \rh - 3\rh^{2} + \frac{6}{5}\rh^{3} - \frac{1}{6}\rh^{4} - \frac{1}{15q^{2}}\right], & 1 \le \ronh < 2; \\
1/r^{2} & \ronh \ge 2. \end{array} \right. \label{eq:cubicsplineforce}
\end{equation}
where $q = r/h$. Integrating a second time using (\ref{eq:Fdr}) gives the kernel used in the gravitational potential, which in this case is given by
\begin{equation}
\phi(r,h) = \left\{ \begin{array}{ll}
1/h \left[\frac{2}{3}\rh^{2} - \frac{3}{10}\rh^4 + \frac{1}{10}\rh^5 - \frac{7}{5} \right], & 0 \le \ronh < 1; \\
1/h \left[\frac{4}{3}\rh^2 - \rh^{3} + \frac{3}{10}\rh^{4} - \frac{1}{30}\rh^{5} - \frac{8}{5} + \frac{1}{15q} \right], & 1 \le \ronh < 2; \\
-1/r & \ronh \ge 2. \end{array} \right. \label{eq:cubicsplinepotential}
\end{equation}

The derivative of the potential with respect to $h$ is given by
\begin{equation}
\pder{\phi}{h} = \left\{ \begin{array}{ll}
1/h^{2} \left[-2 \rh^{2} + \frac32 \rh^4 - \frac35\rh^5 + \frac75 \right], & 0 \le \ronh < 1; \\
1/h^{2} \left[-4\rh^{2} + 4\rh^{3} - \frac32\rh^{4} + \frac15\rh^{5} + \frac85 \right], & 1 \le \ronh < 2; \\
0. & \ronh \ge 2. \end{array} \right. \label{eq:cubicsplinedphidh}
\end{equation}
Alternatively $\partial \phi / \partial h$ can be evaluated from the potential and force functions according to
\begin{equation}
\pder{\phi}{h} = -\frac{1}{h^{2}} \left[ K(q) + q K'(q) \right]
\end{equation}
where $K(q) = h\phi$ and $K'(q) = h^{2}\phi'$ are the functional forms of the potential and force kernels.

\section{Adaptive softening length formalism using averaged softening lengths}
\label{sec:hav}
A alternative way of symmetrising the gravitational potential is to use an average of the particle softening lengths. This is similar to the approach taken in the adaptive smoothing/softening length formalism used by \citet{benz90}. The main difference is that we symmetrise the gravitational potential rather than the force and are therefore able to account for the spatial variation of softening length in the equations of motion, leading to the conservation of both momentum and energy. Note that this is only made possible because of the self-consistent relationship between the density and the smoothing length described in \S\ref{sec:settingh}.

Using an average of the softening lengths, the gravitational part of the Lagrangian can be written in the form
\begin{equation}
L_{grav} = -\sum_{b} m_{b} \Phi_{b} = -\frac{G}{2} \sum_{b} \sum_{c} m_{b} m_{c} \phi_{bc} (\bar{h}_{bc})
\label{eq:Lgravhav}
\end{equation}
where $\phi_{bc}$ refers to $\phi(\vert \br_{b} - \br_{c} \vert)$ and $\bar{h}_{bc} = \frac12 (h_{b} + h_{c})$.
It is then a straightforward matter to derive the equations of motion by using (\ref{eq:Lgravhav}) in the Euler-Lagrange equations (\ref{eq:el}). The derivative of (\ref{eq:Lgravhav}) is given by
\begin{equation}
 \pder{L_{grav}}{{\bf r}_{a}} =  -\frac12 \sum_{b} \sum_{c} m_{b} m_{c} \left[ \left.\pder{\phi_{bc}(\bar{h}_{bc})}{r_{bc}}\right\vert_{h} \frac{\br_{b} - \br_{c}}{\vert \br_{b} - \br_{c} \vert}(\delta_{ba} - \delta_{ca}) + \left.\pder{\phi_{bc}(\bar{h}_{bc})}{\bar{h}_{bc}}\right\vert_{r} \frac12 \left( \pder{h_{b}}{\rho_{b}}\pder{\rho_{b}}{\br_{a}} + \pder{h_{c}}{\rho_{c}}\pder{\rho_{c}}{\br_{a}} \right) \right].
\end{equation}
Using the spatial derivative of the density given by (\ref{eq:gradrho}) and a similar expression for $\partial \rho_{c}/ \partial {\bf r}_{a}$ we have
\begin{eqnarray}
 \pder{L_{grav}}{{\bf r}_{a}} & = & -\frac12 \sum_{b} \sum_{c} m_{b} m_{c} \left.\pder{\phi_{bc}(\bar{h}_{bc})}{r_{bc}}\right\vert_{h} \frac{\br_{b} - \br_{c}}{\vert \br_{b} - \br_{c} \vert}(\delta_{ba} - \delta_{ca})
 \nonumber \\
&  & -\frac12 \sum_{b}\sum_{c}\sum_{d} m_{b} m_{c} m_{d} 
\left.\left.\pder{\phi_{bc}}{\bar{h}_{bc}}\right\vert_{r} \frac12 \left( \pder{h_{b}}{\rho_{b}} \frac{1}{\Omega_{b}} \pder{W_{bd}(h_{b})}{{\bf r}_{a}} \left( \delta_{ba} - \delta_{ca}\right)+ \pder{h_{c}}{\rho_{c}}\frac{1}{\Omega_{c}} \pder{W_{cd}(h_{c})}{{\bf r}_{a}} \left( \delta_{ca} - \delta_{da}\right)\right) 
 \right],
\end{eqnarray}
Collecting terms and simplifying, this expression can be written in the form
\begin{equation}
 \pder{L_{grav}}{{\bf r}_{a}} = -m_{a} \sum_{b} m_{b} \phi'_{ab} \frac{\br_{a} - \br_{b}}{\vert \br_{a} - \br_{b}\vert} - m_{a} \sum_{b} m_{b} \frac12 \left(\frac{\bar{\zeta}_{a}}{\Omega_{a}}  \pder{W_{ab}(h_{a})}{{\bf r}_{a}} + \frac{\bar{\zeta}_{b}}{\Omega_{b}}  \pder{W_{ab}(h_{b})}{{\bf r}_{a}} \right),
\end{equation}
giving the $N-$body equations of motion in the form
\begin{equation}
\frac{d{\bf v}_{a}}{dt}  = -G\sum_{b} m_{b}  \phi'_{ab}(\bar{h}_{ab}) \frac{\br_{a} - \br_{b}}{\vert \br_{a} - \br_{b} \vert}  -\frac{G}{2} \sum_{b} m_{b} \left[ \frac{\bar{\zeta}_{a}}{\Omega_{a}} \pder{W_{ab} (h_{a})}{\br_{a}} +  \frac{\bar{\zeta}_{b}}{\Omega_{b}} \pder{W_{ab} (h_{b})}{\br_{a}}\right], \label{eq:fgradsofthav}
\end{equation}
where in this case we define the quantity $\bar{\zeta}$ according to
\begin{equation}
\bar{\zeta}_{a} \equiv \pder{h_{a}}{\rho_{a}} \sum_{b} m_{b} \pder{\phi_{ab}}{h}(\bar{h}_{ab})
\label{eq:gradsofthav}
\end{equation}
This term is again easily calculated alongside $\rho$ and $\Omega$ during the density summation. However this formalism is quite inefficient in general, since the average $h$ is \emph{only} used in the calculation of $\bar{\zeta}$. The density, SPH and gravity forces are naturally symmetrised by the formulation from a Lagrangian. Calculation of $\bar{\zeta}$ in this case would require an extra loop over the particles. This is for the reason that, whilst the density and smoothing length can be iteratively found for a single particle $a$ (depending only on $h_{a}$), quantities depending on an average smoothing length must be updated when a \emph{neighbouring} value of $h$ has changed (ie. $h_{b}$), leading to a rather inefficient scheme.

 It is worth noting that \citet{hb90} suggested using a Lagrangian to derive an energy-conserving adaptive softening length formalism using an average of the softening lengths some time ago. However contrary to their assertion that `the terms involving $\nabla \epsilon$ will, in general, lead to a violation of linear and angular momentum conservation', the force expressed by (\ref{eq:fgradsofthav}) clearly conserves linear momentum as the summations are antisymmetric in $a$ and $b$. It is also straightforward to show that angular momentum is conserved exactly.

\end{onecolumn}

\label{lastpage}
\end{document}